%% file: main.tex
\documentclass[sigconf]{acmart}

\AtBeginDocument{%
  }

\renewcommand\footnotetextcopyrightpermission[1]{}
\usepackage{booktabs}
\usepackage{amsmath}
\usepackage{graphicx}
\usepackage{xspace}
\usepackage{multirow}
\usepackage{subcaption}
\usepackage{enumitem}
\usepackage[capitalize,noabbrev]{cleveref}
\usepackage{tikz}
\usetikzlibrary{arrows.meta,positioning,calc,shapes.geometric,fit}

\newcommand{\method}{PAIR\xspace}

\newcommand{\eg}{\textit{e.g.},\xspace}
\newcommand{\etal}{\textit{et al.}\xspace}
\newcommand{\boldparagraph}[1]{\par\noindent\textbf{#1:}}

\begin{document}

\title{PAIR: Preference-Aligned Intent Retrieval for Premium People Recommendation at LinkedIn}

\author{Zhongkai Shangguan}
\affiliation{\institution{LinkedIn}\city{Sunnyvale}\state{CA}\country{USA}}
\email{zshangguan@linkedin.com}

\author{Yanbin Kang}
\affiliation{\institution{LinkedIn}\city{Sunnyvale}\state{CA}\country{USA}}
\email{ybkang@linkedin.com}

\author{Yihan Cao}
\affiliation{\institution{LinkedIn}\city{Sunnyvale}\state{CA}\country{USA}}
\email{yihcao@linkedin.com}

\author{Ruijie Jiang}
\affiliation{\institution{LinkedIn}\city{Sunnyvale}\state{CA}\country{USA}}
\email{rjiang@linkedin.com}

\author{Qiannan Yin}
\affiliation{\institution{LinkedIn}\city{Sunnyvale}\state{CA}\country{USA}}
\email{qyin@linkedin.com}

\author{Chang Yue}
\affiliation{\institution{LinkedIn}\city{Sunnyvale}\state{CA}\country{USA}}
\email{cyue@linkedin.com}

\author{Zijian Zhou}
\affiliation{\institution{LinkedIn}\city{Sunnyvale}\state{CA}\country{USA}}
\email{zijzhou@linkedin.com}

\author{Girish Kathalagiri}
\affiliation{\institution{LinkedIn}\city{Sunnyvale}\state{CA}\country{USA}}
\email{gkathalagiri@linkedin.com}

\author{Changshuai Wei }
\affiliation{\institution{LinkedIn}\city{Sunnyvale}\state{CA}\country{USA}}
\email{chawei@linkedin.com}

\author{Zhengming Xing}
\affiliation{\institution{LinkedIn}\city{Sunnyvale}\state{CA}\country{USA}}
\email{zhxing@linkedin.com}


\input{abstract}
\input{intro}
\input{related_v2}
\input{system}
\input{method_v3}
\input{experiments_v3}
\input{conclusion}

\bibliographystyle{ACM-Reference-Format}
\bibliography{references}

\end{document}

%% file: abstract.tex
\begin{abstract}
General-purpose embeddings trained on profile similarity often misrank candidates in professional networks, where relevance is inherently \textit{complementary} rather than \textit{symmetric}. For example, job seekers usually engage with recruiters rather than peers with similar credentials.
Current user recommendation systems are often built on profile similarity or social signals such as shared connections, which are effective in friend-oriented social networks where users tend to connect with similar people or within existing social circles. However, this does not hold in professional platforms like LinkedIn, where interactions are often \textit{goal-driven} and \textit{cross-role interactions}, requiring matching based on complementary intents rather than similarity or social proximity.
In this paper, we propose \textbf{Preference-Aligned Intent Retrieval (\method)}, a retrieval framework that is designed to model user intent preference instead of profile similarity. Our approach is inspired by the preference-learning perspective of Direct Preference Optimization, but instantiates preference learning as an exposure-conditioned multi-negative contrastive objective over embeddings. This formulation simplifies training and scales efficiently to large candidate sets without requiring an explicit reference model.
Furthermore, we incorporate user interaction history by aggregating click histories with positional encoding to capture temporal order, which enables \textit{constant-cost} updates of user representations over a fixed history window, allowing new interactions to be incorporated without re-encoding history. As a result, personalization can be updated efficiently using lightweight CPU operations without repeated LLM inference during serving.
We evaluate \method on LinkedIn’s Premium retrieval task. A 0.6B-parameter model achieves a MAR@100 of 0.45. The model is deployed in LinkedIn Premium Browsemaps, where online A/B tests show improvements in click-through rate and Premium retention without measurable latency regression.
\end{abstract}

\maketitle

%% file: intro.tex
\section{Introduction}
\label{sec:intro}

When a software engineer begins searching for a new role on LinkedIn, their browsing behavior shifts: instead of engaging with fellow engineers, they click on recruiters, hiring managers, and talent acquisition professionals. This shift illustrates a fundamental mismatch in professional network retrieval---relevance is often \textit{complementary} rather than \textit{symmetric}. General-purpose embedding models trained on profile similarity cluster users with similar backgrounds together, but the candidates a job-seeking engineer actually engages with look nothing like the engineer's own profile.

This mismatch directly affects LinkedIn Premium Browsemaps, a people recommendation product for Premium subscribers. Premium Browsemaps surfaces a curated set of profiles on member profile pages, generating profile visits that feed downstream retention signals such as ``\textit{Who viewed your profile}'' notifications. The retrieval stage defines the entire candidate space for downstream ranking: intent-aligned candidates excluded at retrieval cannot be recovered by the ranker. In our offline evaluation, LinkedIn's production member embedding (Holistic-171) achieves MAR@100 of only 0.288 on Premium user retrieval, and a general-purpose off-the-shelf 0.6B LLM embedding scores even lower at 0.192. These results suggest that neither general-purpose collaborative embeddings nor generic semantic embeddings adequately model the complementary and goal-driven interactions underlying professional networking.

Deploying intent-aware retrieval at LinkedIn scale requires addressing three practical constraints. First, the training signal must capture \textit{preference} rather than similarity: clicked candidates should rank above comparable non-clicked candidates, even when the non-clicked candidates are more semantically similar to the query user. Second, user intent is inherently \textit{dynamic}. A member actively job-searching this week may shift toward networking or business development next month. However, repeatedly re-encoding hundreds of millions of user profiles through an LLM whenever interaction history changes is computationally infeasible in production. Third, different recommendation surfaces operate under different latency, memory, and storage constraints, requiring \textit{flexible embedding dimensionality} that can adapt to multiple serving budgets within a single model.

We present \textbf{Preference-Aligned Intent Retrieval (\method)}, a production-oriented retrieval framework for professional people recommendation. \method formulates embedding training as \textit{exposure-conditioned preference learning}: instead of treating negatives as arbitrary in-batch samples, it constructs preference supervision from production exposure logs, where clicked candidates are preferred over exposed-but-unclicked alternatives. The resulting optimization resembles multi-negative contrastive learning, but differs fundamentally in supervision semantics: the production retriever defines an implicit exposure distribution, while user clicks define preferences within that distribution.
To model dynamic intent without introducing online LLM re-inference, \method adopts a triple-tower architecture that combines a static profile embedding with a positional-attention history embedding aggregated from recently clicked profiles. Because history embeddings are precomputed, updating user intent requires only O(Kd) vector operations over a fixed history window. Finally, Matryoshka training~\cite{kusupati2022matryoshka} enables a single model to produce embeddings that can be truncated to different dimensions for different latency and storage budgets across recommendation surfaces.

A 0.6B-parameter \method model achieves MAR@100 of 0.450 on our Premium retrieval benchmark---a 56\% relative improvement over Holistic-171 and 2.3$\times$ over the off-the-shelf base encoder---demonstrating that task-specific preference supervision matters more than model scale for this retrieval task. In an online A/B test on Premium Browsemaps, \method improves click-through rate by 7.5\% and Premium retention rate by 2.1\% over the production baseline, with no measurable latency regression.

\boldparagraph{Contributions}
\begin{itemize}[leftmargin=*]

    \item We formulate Premium people retrieval as \textit{exposure-conditioned preference learning}, where the production retriever defines the exposure distribution and user clicks induce preference supervision. This reframes retrieval optimization from profile similarity matching toward professional intent alignment.

    \item We introduce \method, a production-oriented triple-tower retrieval architecture with additive history fusion and positional attention, enabling dynamic intent updates through \(O(Kd)\) vector operations without online LLM re-inference. Unlike concatenation-based or sequential approaches, history updates require only lightweight embedding-space computation.

    \item We provide a systematic analysis of training and serving tradeoffs for LLM-based retrieval at scale, including mixed-precision stability, temperature sensitivity, history aggregation strategies, and Matryoshka dimensionality across different latency and storage budgets.

    \item We deploy \method in LinkedIn Premium Browsemaps and validate it through both offline retrieval evaluation and an online A/B test, demonstrating improvements in user engagement and Premium retention without measurable serving latency regression.

\end{itemize}

The remainder of this paper is organized as follows. \cref{sec:problem} describes the production system and problem formulation. \cref{sec:method} presents the \method framework. \cref{sec:experiments} reports offline retrieval evaluation and online A/B test results.

%% file: related_v2.tex
\section{Related Work}
\label{sec:related}

\boldparagraph{LLM-Based Embeddings for Retrieval}
Large language models have become effective backbones for dense retrieval. Contrastive pre-training methods such as E5~\cite{wang2022e5} demonstrated that weakly-supervised text pairs can produce competitive embeddings, and subsequent work on synthetic-data-driven training~\cite{wang2024improving} extended this approach to LLM-scale encoders. Recent models further improve embedding quality through architectural and training advances: LLM2Vec~\cite{behnamghader2024llm2vec} converts decoder-only LLMs into bidirectional encoders through masked next-token prediction and contrastive learning, while NV-Embed~\cite{lee2024nvembed} introduces latent attention pooling with two-stage contrastive instruction tuning. These advances establish that LLMs fine-tuned for embedding tasks consistently outperform smaller purpose-built encoders on retrieval benchmarks.

Within recommendation, LLMs have been used as generative recommenders~\cite{geng2022p5,li2023e4srec}, knowledge augmentation modules~\cite{ren2024rlmrec,li2023kar,zhang2024notellm}, and bridges between collaborative filtering and language representations~\cite{li2023ctrl,zhu2024cllm4rec}. At LinkedIn, LLaMA-based dual encoders have been applied to feed retrieval~\cite{ramanujam2025linkedin_feed_llm}, and MixLM~\cite{li2025mixlm} combines embedding and text tokens for efficient reranking. A comprehensive survey by Lin \etal~\cite{lin2025llm_recsys_survey} categorizes these approaches by role.

Our work uses an LLM as the embedding backbone but targets a challenge largely unaddressed by existing methods: reshaping a general-purpose embedding space toward preference-aligned professional retrieval, where relevant candidates are often complementary rather than semantically similar to the query user.

\boldparagraph{Embedding-Based Retrieval at Production Scale}
Dual-encoder retrieval architectures underpin candidate generation in most large-scale recommendation systems. Dense Passage Retrieval (DPR)~\cite{karpukhin2020dpr} established the dual-encoder paradigm for open-domain question answering, and Huang \etal~\cite{huang2020ebr_facebook} scaled embedding-based retrieval (EBR) to Facebook Search with a unified embedding framework, hard negative mining, and ANN-based serving. Yi \etal~\cite{yi2019sampling_bias} addressed sampling bias in two-tower models at YouTube through streaming frequency estimation, showing that correcting for popularity bias in negative sampling substantially improves retrieval quality. Most production retrieval systems optimize semantic or collaborative similarity, which is often misaligned with professional relevance in goal-driven networking settings. In people recommendation specifically, prior work has studied retrieval architectures for employment platforms~\cite{shen2024learning_to_retrieve_job}, while LinkedIn's production systems highlight the importance of scalable retrieval and personalization constraints.

Our setting differs in that professional relevance is often asymmetric and goal-driven: a job seeker engages with recruiters, not with other job seekers. General-purpose member embeddings trained on broad objectives do not capture this complementary intent, motivating a task-specific preference-aligned embedding.

\boldparagraph{Preference Learning from Logged Feedback}
Learning from implicit feedback is central to recommendation, but logged behavior is shaped by the production retrieval policy: clicks are only observed for candidates that were surfaced. Counterfactual learning-to-rank and debiased methods address this through propensity modeling. Direct Preference Optimization (DPO)~\cite{rafailov2023dpo} optimizes language model policies from pairwise preferences without an explicit reward model, and recent work adapts preference optimization to recommendation~\cite{chen2024softmax_dpo}. These methods typically operate on token-level probabilities or generative recommendation policies.

The construction of informative negatives is closely related. ANCE~\cite{xiong2021ance} uses an ANN index to mine hard negatives during training, and Robinson \etal~\cite{robinson2021hard_negatives} provide theoretical analysis showing that harder negatives improve the quality of learned representations in contrastive settings. Our work is inspired by the preference-learning perspective but instantiates it as contrastive retrieval training over embeddings. Rather than mining negatives through ANN search, \method derives preference supervision from production exposure logs: exposed-but-unclicked candidates serve as exposure-conditioned negatives that have already passed the production retrieval filter, while random negatives provide broader coverage. This eliminates the need for a separate reference-model forward pass and naturally encodes the production retriever's exposure distribution.

\boldparagraph{User History Modeling in Retrieval}
Interaction history is a key signal for personalized retrieval. Sequential models such as HSTU~\cite{zhai2024hstu} and its extensions~\cite{dong2025hstu_cp,hertel2024amortized} demonstrate the value of scaling sequence-based user modeling, while multi-interest approaches~\cite{meng2025multi_interest} and large-scale user representations~\cite{zhang2023scaling_user_meta} show that users maintain multiple concurrent intents that a single static profile vector cannot capture. PinnerFormer~\cite{pancha2022pinnerformer} addresses the gap between batch and real-time user representations at Pinterest by training with a dense all-action loss that approximates real-time sequence inference in a batch setting. Beyond recommendation, goal-conditioned and time-aware interactive systems have also demonstrated the importance of adapting model outputs to evolving user goals and temporal context~\cite{shangguan2026advancing,huang2022assister}. Such findings further motivate representations that distinguish relatively stable user characteristics from transient behavioral intent.
However, many history-aware approaches require sequence-model inference or nontrivial state updates when interactions change. Such approaches become prohibitively expensive at LinkedIn scale, where user interaction histories evolve continuously. Our additive history design operates entirely in embedding space: the query representation is the sum of a static profile embedding and a dynamically aggregated history embedding, enabling constant-cost updates over a fixed history window without online LLM re-inference in the serving path. This trades off some sequence modeling expressiveness for predictable, low-cost serving behavior suited to daily or near-real-time intent refreshes.

\boldparagraph{Flexible Embedding Representations}
Embedding dimensionality directly affects storage, latency, and memory bandwidth in large-scale retrieval. Matryoshka Representation Learning~\cite{kusupati2022matryoshka} trains embeddings that remain effective when truncated to lower dimensions. Other approaches explore sparse compression~\cite{kasalicky2025sparse}, analyze embedding collapse under scaling~\cite{guo2023embedding_collapse}, or replace continuous embeddings with discrete semantic IDs~\cite{zheng2025semantic_id,mei2025semantic_ids_music}. We adopt Matryoshka loss within \method so that a single model supports multiple serving configurations, from full 1024-dimensional embeddings for quality-sensitive surfaces to truncated representations for latency-constrained channels, without retraining separate models.

%% file: system.tex
\section{System Overview and Problem Formulation}
\label{sec:problem}

\subsection{Premium Browsemaps Product Surface}
\textit{Premium Browsemaps} is LinkedIn’s people recommendation product tailored for Premium subscribers, designed to enhance member visibility and reinforce the value proposition of LinkedIn Premium through profile discovery and social exposure. 
Surfaced directly on member profile pages, the module recommends a select set of Premium profiles for users to explore. 
Unlike traditional social recommendation systems that primarily optimize for engagement or social proximity, Browsemaps aims to facilitate high-value professional discovery while enabling broader exposure opportunities across Premium members.
From a product perspective, profile visits generated by Browsemaps create a crucial downstream feedback loop tied to Premium retention. When recommended profiles receive traffic, LinkedIn leverages visibility-related signals, such as \textit{"Who viewed your profile"} notifications, to strengthen the user's perception that Premium membership accelerates discoverability and professional reach. 

At production scale, Browsemaps operates as a two-stage retrieval-ranking pipeline serving hundreds of millions of users. In the candidate generation stage, lightweight attribute filters (\eg, location, industry, seniority, and language) first narrow the Premium member pool into a manageable candidate set. An embedding-based retrieval (EBR) system then retrieves the top $\sim 100$ candidates. This retrieval stage is critical, as it defines the candidate space for downstream ranking models and directly shapes both recommendation quality and ecosystem-level exposure allocation.
In the ranking stage, a downstream ranking model scores the retrieved candidates using dense features and selects the final four recommendations displayed to the user.
This work focuses specifically on the retrieval stage, where we demonstrate how preference-aligned representation learning can better capture professional discovery intent beyond simple profile similarity, thereby improving both short-term engagement metrics and long-term ecosystem health.

\subsection{Retrieval Problem in Professional Discovery}

A key challenge in Premium Browsemaps is that professional discovery is often driven by complementary intent rather than symmetric profile similarity. General-purpose embedding retrieval models tend to place users with similar profile attributes close together in the embedding space. However, high-value professional interactions often occur across roles: job seekers may engage with recruiters or hiring managers, business developers may seek potential clients or partners, and talent acquisition professionals may search for candidates with complementary skills. As a result, similarity-based EBR may retrieve professionally similar members while missing candidates that better match the query user's current discovery intent.
This mismatch is particularly important in a two-stage retrieval-ranking system. Since downstream ranking models can only score candidates surfaced by retrieval, intent-aligned candidates excluded during candidate generation cannot be recovered later. We therefore formulate retrieval as a preference-aligned ranking problem: given a query user $q$ and a candidate pool $\mathcal{D}$, the goal is to learn a scoring function $s(q,d)$ such that preferred candidates receive higher retrieval scores than non-preferred alternatives:
\[
s(q,d^+) > s(q,d^-),
\]
where $d^+$ denotes a clicked or positively engaged candidate and $d^-$ denotes an unclicked or sampled negative candidate. This preference-oriented formulation motivates the representation learning framework introduced in \cref{sec:method_pair}.

\subsection{Modeling Static Identity and Dynamic Intent}
\label{sec:features}

Premium subscribers differ from regular LinkedIn users in ways that directly affect retrieval design. They exhibit higher session frequency, stronger goal orientation, and more diverse engagement patterns. Premium users also fall into distinct intent clusters, including \emph{job seekers} browsing recruiters and hiring managers, \emph{business developers} seeking potential clients or partners, and \emph{talent acquisition professionals} sourcing candidates. These intent signals are transient: a user in active job-search mode this week may shift to passive networking next month. Such behavioral dynamics are not captured by static profile attributes alone.
To address this limitation, we represent each query user using two complementary sources of information: static profile identity and dynamic behavioral intent.

\boldparagraph{Profile Features}
Each user profile is serialized into a structured text sequence consumed by the LLM encoder. The sequence includes headline, summary, position history (title, company, duration), education (degree, institution, field), skills, certifications, patents, honors, publications, and languages. Fields are concatenated using section markers and truncated to the encoder's maximum context length. Users without populated profile fields are excluded from the candidate pool.

\boldparagraph{Behavioral Features}
User intent is modeled through click history consisting of the $K$ most recently clicked profiles, ordered by recency. Each clicked profile is represented by a precomputed embedding obtained from the same LLM encoder applied to the clicked user's profile text. The aggregation of these history embeddings into a single intent vector is described in \cref{sec:method_history}. This design separates static identity (who the user is) from dynamic intent (what they currently need), allowing intent representations to be updated without re-encoding the user's own profile.

\subsection{Data Construction}
We construct our training dataset using historical production logs from Premium Browsemaps. For each recommendation impression session, a member's click on a recommended candidate is treated as a positive preference signal ($x^+$). To robustly learn intent-aligned embeddings for retrieval, we construct a contrastive negative candidate pool ($X^-$) for each session using a 1:3 positive-to-negative ratio. This pool consists of two components: (1) \textit{hard negatives}, which are candidates that received impressions in the same session but were not clicked, and (2) \textit{easy negatives}, which are randomly sampled from the active Premium member pool outside the exposed candidate set. Exposed-but-unclicked candidates are prioritized; when fewer than three are available in a session, the remaining slots are filled with random negatives.
Because the logged interactions are collected from the production environment, the observed data is inherently affected by \textit{selection bias} conditioned on the historically deployed retrieval policy. Rather than treating these observational clicks as absolute, static relevance labels, our framework interprets them as \textit{relative preference signals} within the context of the displayed candidate set. This framing aligns naturally with our \method formulation (\cref{sec:method_pair}), where the production model's exposure distribution serves as the implicit reference policy from which preference pairs are constructed. We use a time-based split to ensure that training interactions precede test interactions.

%% file: method_v3.tex
\section{Method}
\label{sec:method}

The production pipeline described in \cref{sec:problem} imposes three practical constraints on retrieval embedding learning: preference alignment beyond profile similarity, serving-efficient history updates without LLM re-inference, and flexible embedding dimensionality across products with different latency and storage budgets. We address these constraints with Preference-Aligned Intent Retrieval (PAIR), a triple-tower intent architecture, additive history aggregation, and Matryoshka training. Our focus is not architectural novelty alone, but retrieval learning under large-scale production serving constraints.

\subsection{Preference-Aligned Intent Retrieval}
\label{sec:method_pair}

PAIR is designed to learn retrieval embeddings from relative user preferences observed in production exposure logs. Initial fine-tuning with standard contrastive learning improved retrieval accuracy over the off-the-shelf baseline, but still produced a similarity-oriented embedding space: the model tended to cluster users with similar profiles, while many high-value professional interactions are complementary. For example, a job-seeking engineer may be matched with other engineers, even though their observed behavior indicates stronger interest in recruiters or hiring managers.

To address this mismatch, we formulate PAIR as an exposure-conditioned preference contrastive learning protocol. The objective is conceptually inspired by relative preference learning, but it is not a direct embedding-space instantiation of Direct Preference Optimization (DPO). In particular, PAIR does not explicitly model a reference policy or optimize a token-level probability ratio. Instead, the historically deployed retriever defines the exposure distribution from which preference comparisons are observed.

Given a query user $q$, one clicked candidate $d^+$, and $K$ negative candidates $\{d_i^-\}_{i=1}^{K}$, we define the retrieval score as:
\begin{equation}
    s(q,d) = \cos(\mathbf{e}_q^{\mathrm{final}}, \mathbf{e}_d) / \tau,
\end{equation}
where $\mathbf{e}_q^{\mathrm{final}}$ is the final query representation, $\mathbf{e}_d$ is the candidate embedding, and $\tau$ is a temperature parameter. We optimize the following multi-negative objective:
\begin{equation}
\mathcal{L}_{\text{PAIR}} = -\log
\frac{\exp(s(q,d^+))}
{\exp(s(q,d^+)) + \sum_{i=1}^{K}\exp(s(q,d_i^-))}.
\label{eq:pair}
\end{equation}
Mathematically, \cref{eq:pair} has the same computational form as multi-negative contrastive learning~\cite{oord2018infonce}. The distinction of PAIR lies in the semantics of its supervision. Unlike conventional hard-negative mining, exposed-but-unclicked candidates are not merely close neighbors in embedding space. They are candidates selected by the deployed retrieval policy and therefore represent realistic alternatives that the production system would actually consider. PAIR therefore optimizes preference ordering within the support of the serving policy, rather than only separating arbitrary or ANN-mined negatives.

Specifically, positives and negatives are constructed from production exposure logs: clicked candidates are treated as preferred, exposed-but-unclicked candidates are treated as exposure-conditioned negatives, and randomly sampled candidates provide broader coverage of the member space. Thus, the model learns exposure-conditioned professional preferences rather than profile similarity alone.

This design makes the training protocol compatible with production-scale embedding retrieval because it avoids an explicit reference-model forward pass. However, because preferences are only observed within exposed candidate sets, PAIR inherits selection bias from the historical retrieval policy. We view this as a practical production tradeoff: the objective improves retrieval quality under realistic exposure distributions, while future iterations can incorporate exploration-aware sampling to mitigate feedback loops.

\subsection{Triple-Tower Intent Architecture}
\label{sec:method_arch}

\begin{figure*}[t]
    \centering
    \includegraphics[width=0.8\textwidth]{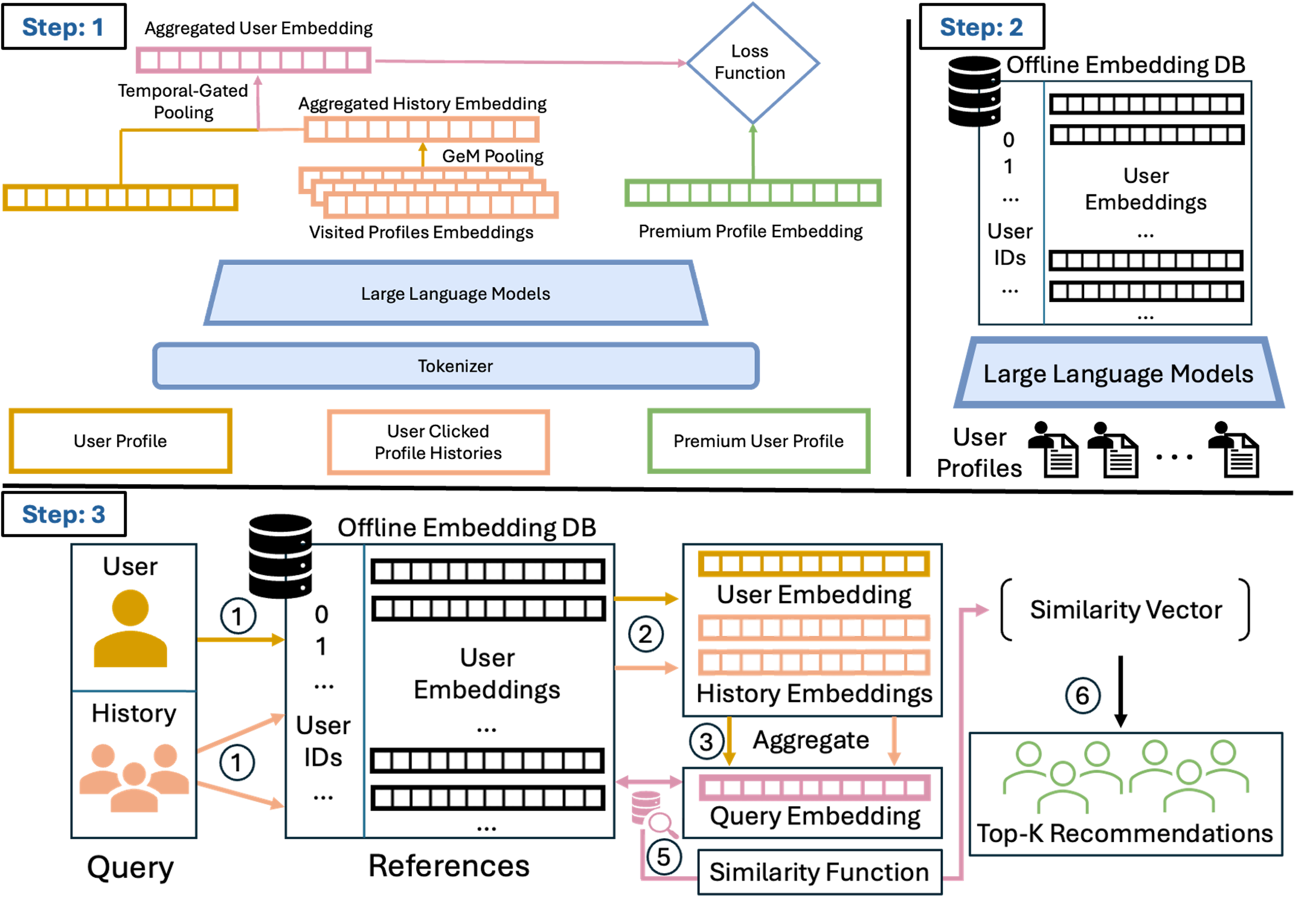}
    \caption{ \textit{Preference-Aligned Intent Retrieval architecture.}
    The model adopts a triple-tower design. 
    The profile tower and document tower share parameters to encode query and candidate profiles. 
    The history tower aggregates recent interaction embeddings using positional attention to capture short-term behavioral intent. 
    The final query representation combines static profile identity and dynamic intent through normalized additive fusion, enabling efficient history updates without re-encoding the full profile through the LLM.
    }
    \label{fig:architecture}
\end{figure*}

To separate stable profile identity from transient behavioral intent, we adopt a triple-tower architecture, as shown in \cref{fig:architecture}. The query profile tower encodes the query user's serialized profile text as $\mathbf{e}_q=f_\theta(x_q)$, the document tower encodes candidate profiles as $\mathbf{e}_d=f_\theta(x_d)$, and the history tower aggregates embeddings of recently clicked profiles into an intent vector $\mathbf{e}_{\mathcal{H}}$. The query profile tower and document tower share the same encoder parameters, while the history tower operates on precomputed profile embeddings.

A direct unnormalized addition of profile and history embeddings may lead to scale mismatch between towers. We therefore use normalized additive fusion:
\begin{equation}
    \tilde{\mathbf{e}}_q =
    \frac{\mathbf{e}_q}{\|\mathbf{e}_q\|_2},
    \quad
    \tilde{\mathbf{e}}_{\mathcal{H}} =
    \frac{\mathbf{e}_{\mathcal{H}}}{\|\mathbf{e}_{\mathcal{H}}\|_2},
\end{equation}
\begin{equation}
    \mathbf{e}_q^{\mathrm{final}} =
    \mathrm{L2Norm}\left(
    \tilde{\mathbf{e}}_q + \gamma \tilde{\mathbf{e}}_{\mathcal{H}}
    \right),
\label{eq:additive}
\end{equation}
where $\gamma$ is a learnable scalar that controls the contribution of behavioral history. Candidate embeddings are also L2-normalized before cosine retrieval. This normalization prevents either the profile tower or the history tower from dominating the final representation due to norm differences.

The additive design allows profile embeddings to be computed offline, while intent updates require only lightweight vector operations when user histories change. We use Qwen3-Embedding-0.6B as the base encoder, a 0.6B-parameter model selected to balance retrieval quality and batch inference throughput. In our production setting, it can process active Premium users within the daily batch window, while larger models introduce substantially higher GPU cost without proportional retrieval-quality gains.

\hspace{-2cm}
\subsection{Additive History with Positional Attention}
\label{sec:method_history}

Given the $K$ most recently clicked profiles $\mathcal{H}_q=\{h_1,\ldots,h_K\}$, each clicked profile is represented by a precomputed embedding $\mathbf{e}_{h_i}=f_\theta(x_{h_i})$. Because recent clicks often better reflect current user intent, we compute a position-aware attention weight:
\begin{equation}
    \alpha_i =
    \frac{
    \exp\!\left(\mathbf{w}^{T}(\mathbf{e}_{h_i}+\mathrm{PE}(i))\right)
    }
    {
    \sum_{j=1}^{K}
    \exp\!\left(\mathbf{w}^{T}(\mathbf{e}_{h_j}+\mathrm{PE}(j))\right)
    },
    \quad
    \mathbf{e}_{\mathcal{H}} = \sum_{i=1}^{K}\alpha_i \mathbf{e}_{h_i},
\label{eq:patt}
\end{equation}
where $\mathrm{PE}(i)$ is a learned positional encoding for position $i$, with the most recent click assigned to $i=1$, and $\mathbf{w}$ is a learnable attention vector. Both are trained end-to-end with the PAIR objective.

The attention module is intentionally query-agnostic. This design learns a global recency- and item-importance weighting over the history window, allowing the history vector to be refreshed independently of the query profile tower. Query-conditioned attention is more expressive, since it can model interactions between the current profile representation and each historical click. However, it introduces additional query-history interaction computation and complicates serving. In contrast, our query-agnostic design preserves the production property that history updates require only lightweight vector operations and no query-history joint inference.

With a fixed history window, updating $\mathbf{e}_{\mathcal{H}}$ requires only $O(Kd)$ vector operations and no LLM re-inference. This is significantly cheaper than concatenating profile and history text followed by full LLM encoding. We set $K{=}10$ to cover the typical recent interaction window of active Premium users while keeping serving cost bounded. The design intentionally trades off some sequence modeling expressiveness for predictable, low-cost serving behavior, making it suitable for hourly or near-real-time intent refreshes in production. We evaluate this trade-off through pooling-based history aggregation ablations in \cref{sec:experiments}.

\subsection{Dimension-Flexible Embedding Training}
\label{sec:method_matryoshka}

Different LinkedIn surfaces operate under different latency and storage budgets. Premium Browsemaps can use full 1024-dimensional embeddings to maximize retrieval quality, while latency-sensitive products may require smaller representations such as 256 dimensions. Training separate models for each serving configuration would increase maintenance costs and complicate deployment.

We therefore adopt Matryoshka loss~\cite{kusupati2022matryoshka}, optimizing PAIR jointly across multiple prefix dimensions 
\(d \in \{200, 256, 512, 1024\}\):
\begin{equation}
    \mathcal{L} = \sum_{d \in \mathcal{D}} \mathcal{L}_{\text{PAIR}}^{(d)},
\label{eq:matryoshka}
\end{equation}
where \(\mathcal{D}=\{200,256,512,1024\}\), and \(\mathcal{L}_{\text{PAIR}}^{(d)}\) denotes the PAIR loss computed using only the first \(d\) embedding dimensions. The 200-dimensional prefix is included to approximate the storage footprint of the production 171-dimensional embedding, while 256, 512, and 1024 dimensions represent progressively higher-quality serving configurations.

At serving time, embeddings can be truncated to the target dimension without additional computation or retraining. This enables a single model to support multiple retrieval surfaces with different efficiency requirements while preserving a shared training and deployment pipeline.

\subsection{Production Serving Considerations}
\label{sec:serving}

The proposed design is optimized for production retrieval rather than only offline evaluation. Profile and candidate embeddings are computed through an offline batch pipeline, while behavioral intent is updated through lightweight vector aggregation over precomputed embeddings. This avoids LLM re-inference when user click histories change and makes the model suitable for large-scale serving.
The Matryoshka training objective further allows the same model to support multiple embedding dimensions under different latency and storage budgets. Premium Browsemaps uses the full embedding representation to maximize retrieval quality, while latency-sensitive surfaces can truncate embeddings at serving time without retraining a separate model.

%% file: experiments_v3.tex
\hspace{-15cm}
\section{Experiments}
\label{sec:experiments}
We evaluate \method along three dimensions. First, we measure offline retrieval quality against LinkedIn's production member embedding and an off-the-shelf 0.6B embedding encoder. Second, we analyze the contribution of history augmentation and the quality-efficiency tradeoff enabled by Matryoshka dimensionality. Third, we validate whether offline retrieval gains translate into online engagement improvements through a controlled A/B test on Premium Browsemaps.

\subsection{Experimental Setup}
\label{sec:exp_setup}

\boldparagraph{Model and training}
The base encoder is an internally trained 0.6B-parameter embedding model based on the Qwen architecture. We train the model with DeepSpeed in bf16 precision on 8 GPUs. In preliminary tuning, fp16 mixed-precision training caused gradient overflow under DeepSpeed, so all reported models use bf16. We set the temperature to $\tau = 0.02$ for all experiments and use cosine learning-rate scheduling with warmup and gradient accumulation.

\boldparagraph{Data}
We use a time-based split, with all training interactions occurring before the test window, to avoid leakage from future user behavior. The training set contains 209K query-level training instances involving 4.4M unique member profiles across queries, candidates, and history users. The test set contains 11.8K query-level instances involving 93K unique member profiles. Each instance consists of one clicked candidate as the positive and three negatives sampled from non-clicked impressions and randomly sampled members. User profiles include headline, summary, positions, education, skills, certifications, patents, honors, publications, and languages. Users with no profile information are filtered.

\boldparagraph{Metrics}
We report two offline retrieval metrics. \textbf{Accuracy (ACC)} measures whether the model assigns the highest similarity score to the clicked candidate among four candidates consisting of one positive and three negatives. The random baseline is 0.25. \textbf{Mean Average Recall at $k$ (MAR@$k$)} evaluates whether clicked candidates are ranked within the top-$k$ positions of a fixed 100-candidate evaluation pool constructed from logged candidates and sampled negatives. The same candidate pool is used for all models to ensure a fair comparison. Since ACC is computed over only four candidates and quickly reaches saturation, MAR@$k$ provides the more informative offline retrieval signal.

\boldparagraph{Baselines}
We compare against three groups of baselines: (1) \textbf{Holistic-171}, LinkedIn's production general-purpose member embedding with 171 dimensions, trained across multiple LinkedIn engagement objectives using both profile and behavioral signals; (2) the off-the-shelf 0.6B embedding model without task-specific fine-tuning; and (3) \method-trained models initialized from the same 0.6B encoder, with and without history augmentation.

\subsection{Main Offline Results}
\label{sec:exp_main}

\begin{table}[t]
\caption{\textbf{Main results on Premium user retrieval.} Off-the-shelf models are evaluated without fine-tuning. \method models are fine-tuned from the 0.6B embedding encoder. ACC is computed over 4 candidates consisting of 1 positive and 3 negatives. MAR@$k$ is evaluated on a fixed pool of 100 candidates.}
\label{tab:main_results}
\centering
\small
\resizebox{\columnwidth}{!}{%
\begin{tabular}{lccc}
\toprule
\textbf{Model} & \textbf{ACC}~$\uparrow$ & \textbf{MAR@25}~$\uparrow$ & \textbf{MAR@100}~$\uparrow$ \\
\midrule
\multicolumn{4}{l}{\textit{Baselines}} \\
Holistic-171 & 0.794 & 0.169 & 0.288 \\
Off-the-shelf 0.6B embedding & 0.771 & 0.100 & 0.192 \\
\midrule
\multicolumn{4}{l}{\textit{\method fine-tuned from the 0.6B encoder}} \\
Two-tower & 0.960 & 0.255 & 0.436 \\
+ Max history  & 0.963 & 0.252 & 0.430 \\
+ Mean history  & 0.965 & 0.253 & 0.430 \\
+ GeM history & 0.966 & 0.264 & 0.447 \\
+ Attention history & 0.965 & 0.246 & 0.431 \\
+ P-ATT history & \textbf{0.967} & \textbf{0.265} & \textbf{0.450} \\
\bottomrule
\end{tabular}%
}
\end{table}

\cref{tab:main_results} presents the main offline retrieval results on the Premium user retrieval benchmark. We compare LinkedIn's production Holistic-171 embedding, the off-the-shelf 0.6B embedding encoder without task-specific fine-tuning, and several \method variants trained with Premium preference supervision. The results allow us to separate the effect of preference fine-tuning from the additional contribution of history augmentation.

\boldparagraph{Preference fine-tuning is the dominant source of improvement}
The \method model with P-ATT history achieves MAR@100 of 0.450, a 56\% relative improvement over the production Holistic-171 embedding and a 2.3$\times$ improvement over the off-the-shelf 0.6B embedding model. Even without history augmentation, the two-tower \method model reaches MAR@100 of 0.436, improving over the production baseline by 51\%. This indicates that adapting the embedding space to Premium click-derived preference data is the dominant source of offline retrieval improvement.

\boldparagraph{History augmentation provides a moderate additional gain}
Adding P-ATT history improves MAR@100 from 0.436 to 0.450, corresponding to a 3.2\% relative improvement over the history-free two-tower model. This gain is smaller than the improvement from preference-based fine-tuning, indicating that the dominant improvement comes from adapting the embedding space to Premium preference supervision. History augmentation therefore serves primarily as a deployment-efficient mechanism for incorporating dynamic intent, yielding additional gains on top of preference fine-tuning while avoiding online LLM re-inference.

\boldparagraph{ACC is close to saturation}
All \method variants achieve ACC above 0.96, compared with the 0.25 random baseline. This suggests that the 4-candidate discrimination task is nearly saturated. We therefore treat MAR@$k$ over the 100-candidate pool as the primary offline metric for comparing retrieval quality.

\boldparagraph{Dimensionality alone does not explain the gain}
Although \method uses a larger default embedding dimension than Holistic-171, the improvement is not solely due to dimensionality. As shown in \cref{tab:matryoshka}, the 256-dimensional version of \method still achieves MAR@100 of 0.432, substantially outperforming the 171-dimensional production baseline.

\subsection{History Augmentation Analysis}
\label{sec:exp_history}

History augmentation captures short-term intent that is not available in static profile features. We compare five aggregation methods over the $K$ most recent clicked profiles: max pooling, mean pooling, GeM pooling, attention pooling, and positional attention. All history methods operate on precomputed profile embeddings and therefore avoid LLM re-inference during history updates.

\begin{table}[t]
\caption{\textbf{History aggregation tradeoff.} All history methods aggregate $K{=}10$ precomputed history embeddings with 1024 dimensions. None require LLM re-inference during history updates.}
\label{tab:pooling_serving}
\centering
\small
\resizebox{\columnwidth}{!}{%
\begin{tabular}{lccl}
\toprule
\textbf{Method} & \textbf{MAR@100}~$\uparrow$ & \textbf{Update Cost} & \textbf{Operations} \\
\midrule
No history & 0.436 & -- & Static profile only \\
Max pooling & 0.430 & $O(Kd)$ & Element-wise max \\
Mean pooling & 0.430 & $O(Kd)$ & Element-wise mean \\
GeM pooling & 0.447 & $O(Kd)$ & Power mean \\
Attention & 0.431 & $O(Kd)$ & Dot product + softmax + weighted sum \\
P-ATT & \textbf{0.450} & $O(Kd)$ & Positional encoding + dot product + softmax + weighted sum \\
\bottomrule
\end{tabular}%
}
\end{table}

As shown in \cref{tab:pooling_serving}, P-ATT achieves the highest MAR@100, followed closely by GeM pooling. The margin between P-ATT and GeM is modest, indicating that much of the benefit comes from incorporating recent behavioral context rather than from a highly expressive sequence model. Some simple aggregation methods underperform the history-free baseline, suggesting that history signals must be incorporated carefully rather than added indiscriminately.

P-ATT provides a simple way to encode recency while preserving the constant-cost update property needed for serving. The attention is intentionally query-agnostic: it learns a global recency- and item-importance weighting over the history window, allowing the history vector to be refreshed independently of the query profile tower. This design is less expressive than query-conditioned attention, but avoids query-history joint inference and keeps history updates lightweight.

\subsection{Dimension and Serving Tradeoff}
\label{sec:exp_tradeoffs}

Embedding dimensionality directly affects storage cost, memory bandwidth, and retrieval latency in large-scale serving. We therefore evaluate the same Matryoshka-trained \method model under multiple truncated prefix dimensions. This experiment focuses on the quality-efficiency tradeoff enabled by dimension-flexible embeddings, rather than comparing separately trained models for each dimension.

\begin{table}[t]
\caption{\textbf{Matryoshka dimension-quality tradeoff.} All variants are produced by a single \method model trained with joint Matryoshka loss. Relative storage is normalized to the 1024-dimensional embedding. Relative latency is based on dimension-dependent retrieval cost and fixed serving overhead.}
\label{tab:matryoshka}
\centering
\small
\resizebox{\columnwidth}{!}{%
\begin{tabular}{lcccl}
\toprule
\textbf{Dim} & \textbf{MAR@100}~$\uparrow$ & \textbf{Rel. Storage} & \textbf{Rel. Latency} & \textbf{Intended Surface} \\
\midrule
200  & 0.421 & 0.20$\times$ & 0.35$\times$ & Candidate pre-filtering \\
256  & 0.432 & 0.25$\times$ & 0.40$\times$ & Latency-sensitive channels \\
512  & 0.444 & 0.50$\times$ & 0.65$\times$ & Standard retrieval \\
1024 & 0.450 & 1.00$\times$ & 1.00$\times$ & Premium Browsemaps \\
\bottomrule
\end{tabular}%
}
\end{table}

As shown in \cref{tab:matryoshka}, full 1024-dimensional embeddings achieve the best retrieval quality, while lower-dimensional prefixes preserve much of the gain at substantially lower storage and latency cost. The 256-dimensional variant achieves MAR@100 of 0.432, outperforming the 171-dimensional production embedding while requiring only 25\% of the storage of the full 1024-dimensional representation. The 200-dimensional prefix is included to approximate the storage footprint of the production 171-dimensional embedding. These results indicate that a single Matryoshka-trained model can support multiple recommendation surfaces with different efficiency requirements without retraining separate models.

\subsection{Online A/B Test}
\label{sec:exp_online}

Offline metrics measure retrieval quality in isolation. Online evaluation is necessary to validate whether improved candidate generation translates into user-facing engagement and business outcomes. We conducted a controlled A/B test on LinkedIn Premium Browsemaps to measure the effect of replacing the production retrieval embedding with \method embeddings in the recommendation pipeline.

\boldparagraph{Experiment setup}
The treatment group receives recommendations retrieved using \method embeddings with P-ATT history augmentation at 1024 dimensions, while the control group uses the production Holistic-171 embedding. Both groups use the same downstream ranker, allowing us to isolate the effect of the retrieval embedding. Traffic is split at the user level and stratified by subscription tier and activity level to ensure balanced cohorts. Guardrail metrics, including page load latency, recommendation coverage, and serving error rate, are monitored continuously to detect potential regressions.

\begin{table}[t]
\caption{\textbf{Online A/B test results on Premium Browsemaps.} Treatment: \method{} with P-ATT history augmentation at 1024 dimensions. Control: production Holistic-171 embedding. Both arms use the same downstream ranker. Metrics are reported as relative changes over the control group. Statistical significance under the production experimentation framework is indicated by $^*$.}
\label{tab:ab_test}
\centering
\small
\begin{tabular}{lc}
\toprule
\textbf{Metric} & \textbf{Relative Change} \\
\midrule
Click-through rate (CTR)   & +7.5\%$^*$ \\
Premium retention rate & +2.1\%$^*$ \\
\midrule
Page load latency          & +0.1\% \phantom{$^*$} \\
Recommendation coverage    & -0.2\% \phantom{$^*$} \\
\bottomrule
\end{tabular}
\end{table}

\boldparagraph{Online results}
As shown in \cref{tab:ab_test}, replacing the production embedding with \method improves online engagement. CTR increases by 7.5\%, indicating that the retrieval stage surfaces candidates that better match user intent before downstream ranking. Premium retention also improves by 2.1\%, suggesting that better profile discovery and increased exposure are associated with improved Premium user outcomes.

Importantly, the treatment does not degrade serving reliability. Page load latency changes by only 0.1\%, and recommendation coverage remains effectively unchanged. These guardrail results suggest that the embedding replacement improves recommendation quality without introducing measurable serving regressions.

\boldparagraph{Discussion}
The online results are directionally consistent with the offline findings in \cref{sec:exp_main}. Although the offline MAR@100 improvement over the production embedding is large, the online effect size is smaller because retrieval changes are filtered through the downstream ranker and user-facing product constraints. This attenuation is expected in a multi-stage recommendation system: candidate generation improves the available candidate set, while final engagement depends on ranking, presentation, and user context.

\boldparagraph{Limitations}
The A/B test evaluates \method within a fixed downstream ranker; joint optimization of retrieval and ranking may yield additional gains. The experiment primarily measures short-term engagement and retention outcomes within the observed experiment window, while longer-term subscription effects require extended observation. Additionally, while the mixed-negative construction used in training reflects a practical production tradeoff, a more exhaustive comparison of random-only, impression-only, and mined hard negatives remains future work. Finally, the test is conducted on Premium Browsemaps only; generalization to other surfaces, such as email recommendations, job matching, or other member recommendation surfaces, remains to be validated.

%% file: conclusion.tex
\section{Conclusion}
\label{sec:conclusion}

We present \method, a production-oriented retrieval framework for professional people recommendation in LinkedIn Premium Browsemaps. The core idea is to learn retrieval embeddings from exposure-conditioned user preferences rather than profile similarity alone. By fine-tuning a 0.6B-parameter embedding model on Premium click-derived preference signals, \method achieves a MAR@100 of 0.450, corresponding to a 56\% relative improvement over the production Holistic-171 embedding and a 2.3$\times$ improvement over the off-the-shelf base encoder. These results show that task-specific preference supervision can substantially reshape a general embedding space for professional discovery.

Beyond preference fine-tuning, we introduce a serving-efficient intent representation that combines static profile identity with dynamic click-history signals. The normalized additive fusion design captures transient user intent while allowing history updates through lightweight $O(Kd)$ vector operations for a fixed history window $K$, avoiding LLM re-inference in the serving path. Matryoshka training further enables a single model to support multiple embedding dimensions, allowing retrieval surfaces with different latency and storage budgets to trade off efficiency and quality without retraining separate models.

\hspace{-15cm}